\documentclass[aps,prl,superscriptaddress,twocolumn,showpacs,amsmath,floatfix,citeautoscript]{revtex4-2}
\usepackage{graphicx}
\usepackage{float}
\usepackage{dcolumn}
\usepackage{color}
\usepackage{latexsym,bm}
\usepackage[normalem]{ulem}
\usepackage{multirow}
\usepackage{appendix}

\usepackage{amsmath}

\begin{document}

\hyphenpenalty=5000
\tolerance=1000

%\title{Time-dependent density functional theory investigation of shift current dynamics in WS$_{2}$ }
\title{Ultrafast shift current dynamics in WS$_{2}$ monolayer}
\author{Fuxiang He}
\affiliation{Key Laboratory of Quantum Information, University of Science and
  Technology of China, Hefei, Anhui, 230026, People's Republic of China}
\affiliation{Synergetic Innovation Center of Quantum Information and Quantum
  Physics, University of Science and Technology of China, Hefei, 230026, People's Republic of China}
 \author{Xinguo Ren}
\email{renxg@iphy.ac.cn}
\affiliation{Beijing National Laboratory for Condensed Matter Physics, Institute of Physics, Chinese Academy of Sciences}
\author{Sheng Meng}
\affiliation{Beijing National Laboratory for Condensed Matter Physics, Institute of Physics, Chinese Academy of Sciences}
\author{Lixin He}
\email{helx@ustc.edu.cn}
\affiliation{Key Laboratory of Quantum Information, University of Science and
  Technology of China, Hefei, Anhui, 230026,  People's Republic of China}
\affiliation{Synergetic Innovation Center of Quantum Information and Quantum
  Physics, University of Science and Technology of China, Hefei, 230026, People's Republic of China}
\date{\today}
%%%%%%%%%%%%%%%%%%%%%%%%%%%%%%%%%%%%%%%%%%%%%%%%%%%%%%%%%%%%%%%%%%%%%%%%%%%%%%%%%%%%%%%%%%%%%%%%%%%%%%%%%%%%%%%%%%%%%%%%%%%%%%%%%%%
%%%%%     Title
%%%%%%%%%%%%%%%%%%%%%%%%%%%%%%%%%%%%%%%%%%%%%%%%%%%%%%%%%%%%%%%%%%%%%%%%%%%%%%%%%%%%%%%%%%%%%%%%%%%%%%%%%%%%%%%%%%%%%%%%%%%%%%%%%%%

\begin{abstract}
The shift current effect, in materials lacking inversion symmetry, may potentially allow the performance of photovoltaics to surpass  the Shockley–Queisser limit for traditional p-n junction-based photovoltaics. Although the shift-current effect has been studied from first-principles via second-order perturbation theory, an understanding of the dynamics of hot carriers is still lacking. We investigate the dynamics of the shift current in monolayer WS$_{2}$ via real-time propagation time-dependent density functional theory (rt-TDDFT). We find that the shift current can be generated within 10 - 20 fs after turning on the lights and dissipates within approximately a few tens of femtoseconds after turning off the lights.  This property can be used for ultrafast photon detection. This work provides an important step toward understanding the dynamics of shift-current effects, which is crucial for device applications.

%Keywords: Azobenzene, Real-time TDDFT, photoisomerization.
\end{abstract}

\maketitle

%\section{Introduction}

In materials lacking inversion symmetry,
current can be generated under illumination, which is known
as the bulk photovoltaic effect (BPVE) or ``shift current'' effect \cite{chynoweth_surface_1956,fridkin_photoconductivity_1974,sturman_photovoltaic_1992}.
The direction of the generated current can be reversed via external voltages as demonstrated in the BiFeO$_3$ single crystal\cite{choi2009switchable}.
The shift current effect (or the BPVE) can be used as an alternative to the photocurrent generated by traditional semiconductor p–n junctions \cite{tan_shift_2016, cook_design_2017}.
More remarkably, it allows the generation of a photovoltage far above the band gap~\cite{glass_highvoltage_1974,dalba_giant_1995,yang_above-bandgap_2010}.
For example, the measured open-circuit voltage in BFO can be as large as 30 V, ten times larger than its band gap of $\sim$ 2.7 eV~\cite{alexe2011tip}.
It therefore may potentially allow the performance of the BPVE-based photovoltaics to surpass the Shockley–Queisser limit for the traditional p-n junction based photovoltaics \cite{spanier_power_2016}.

The shift-current effect is a nonlinear optical response \cite{sturman_photovoltaic_1992,sipe_second-order_2000}.
It is now understood that the shift current originates from the change in the Berry connection of electron bands upon optical transition \cite{morimoto2016topological,Hatada2020}.
Recently the shift-current conductivity has been calculated via first-principles methods\cite{young_first_2012,zheng_first-principles_2015,
wang_first-principles_2017,ibanez-azpiroz_ab_2018, wang_first-principles_2019}, which may provide useful guidance to find suitable materials for PV and other applications\cite{cook_design_2017}.

Despite intensive investigations over the past decades, the understanding of the shift current is far from complete,
especially the theoretical investigation of the dynamics of excited ``hot'' carriers is still lacking.
It has been shown that PV conversion efficiency is closely related to the diffusion
length and carrier relaxation time\cite{zenkevich2014giant,spanier_power_2016}.
To achieve high efficiency, it is crucial to understand the carrier dynamics to optimize PV devices.

In this work, we investigate the shift-current dynamics in  monolayer WS$_{2}$ via  the real-time propagation time-dependent density functional theory (rt-TDDFT)\cite{Theilhaber1992,Castro2004,Meng2008,Lian2018}.
We exam the influence of temperature on the excitation and dissipation of the shift current. We show that the shift-current can be established within 10 - 20 fs after turning on the light and diminishes on a similar time
scale when the light is turned off, which can be used for ultrafast photon detection.
Our work provides an important step to understanding the dynamics of shift currents from first-principles calculations.

%\section{Methods}

%\subsection{Calculation of shift-current via rt-TDDFT method}

The shift current conductivities have been calculated via a second-order perturbation theory\cite{sipe_second-order_2000}.
In this work, we calculate the photonexcited current directly via time-dependent wave functions.
The current density $\boldsymbol{J} (t)$ under the light field is given by
\begin{equation}
\boldsymbol{J}(t)= \frac{e \hbar}{m} \sum_{n=1}^{\rm occ}\sum_{\boldsymbol{k}} \operatorname{Im} \left[ \langle  \tilde{\psi}_{n\boldsymbol{k}}(t)|\nabla| \tilde{\psi}_{n \boldsymbol{k}}(t) \rangle \right]\, ,
\end{equation}
where $\tilde{\psi}_{n \boldsymbol{k}}(t)$ is the time-dependent Bloch wave function of the $n$-th band at point $\boldsymbol{k}$.
The  time-dependent wave functions are obtained by the real-time evolution of the time-dependent Kohn-Sham equation, under
an external time-dependent electric field $\boldsymbol{E}(\omega, t)$.

To be compatible with the periodic boundary conditions, a velocity gauge is used for the electric field\cite{pemmaraju2018velocity}, i.e.,
\begin{equation}
\boldsymbol{E}(\omega, t)=-{1 \over c} {\partial \boldsymbol{A} (\omega, t) \over \partial t} .
\end{equation}
The vector potential is taken to be uniform in the system, i.e.,
\begin{equation}
 \boldsymbol{A} (\omega, t) =  \boldsymbol{A}_0 \cos (\omega t) \, .
\end{equation}

The time-dependent Kohn-Sham equation under the vector potential reads~\cite{pemmaraju2018velocity},
\begin{equation}
i \hbar \frac{\partial}{\partial t} \tilde{\psi}_{i}(t)
=\left\{\frac{1}{2 m}\left[\boldsymbol{p}+\frac{e}{c} \boldsymbol{A}(t)\right]^{2}+\hat{\tilde{V}}^{\rm eff}(t) \right\} \tilde{\psi}_{i}(t) \, ,
\label{eq:tddft}
\end{equation}
where $\hat{\tilde{V}}^{\rm eff}(t) $=$V_{Hxc}[n(t)]$+$\hat{\tilde{V}}_{\rm ion}$, with
$V_{Hxc}[n(t)]$  being the time-dependent Hartree-exchange-correlation potential, and $\hat{\tilde{V}}_{\rm ion}$  the electron-ion interaction
operator under the vector potential \cite{pemmaraju2018velocity}. For simplicity, we neglect the real-space symbol $\boldsymbol{r}$ in the equation.
We solve  Eq. (\ref{eq:tddft}) via real-time evolution of the wave function using the Crank-Nicholson  scheme\cite{Castro2004}.

%\subsection{Calculation details}
The shift current of the WS$_{2}$ monolayer is simulated via the rt-TDDFT methods implemented in the Atomic-orbital Based Ab-initio Computation at USTC (ABACUS) package~\cite{Li2016}.
ABACUS is developed to perform large-scale density functional theory (DFT) calculations based on numerical atomic orbitals (NAO) \cite{Chen2011}.
We adopt the Perdew-Burke-Ernzerhof (PBE) exchange-correlation functional~\cite{perdew1996} and the SG15\cite{Schlipf2015} optimized norm-conserving Vanderbilt (ONCV) pseudopotentials\cite{Hamann2013}
are used to describe the interactions between ions and valence electrons.
The DFT-D3 correction of Grimme \cite{ Grimme2010D3} is used to describe the van der Waals interactions.
A uniform real-space grid corresponding to an energy cutoff of 240 Ry for the charge density is adopted to
solve the Poisson equation via fast Fourier transform.
The 2$s$2$p$1$d$ NAO basis set for S, and the 4$s$2$p$2$d$2$f$   basis set for W  are used to expand the Kohn-Sham wave functions.
The cutoff radii of the NAOs are set to 10 a.u.

We simulate the monolayer WS$_{2}$ supercell containing 4$\times$4$\times$1 primitive cells and
a 17 \AA ~vacuum is added to avoid interaction between the periodic images.
A 24×24×1 $\boldsymbol{k}$-mesh is used to sample the Brillouin zone, which converges the results very well.
The time step for the electron wave function evolution is set to 0.05 fs.
To study the temperature effects and the electron-phonon interactions, we also perform molecular dynamics (MD) for the ions along with the electron wave function evolution.
The calculated currents are averaged for 50 fs, after they are stabilized, which converge the results very well.

%\section{Results and Discussion}

\begin{figure}
  \centering
  \includegraphics[width=0.48\textwidth]{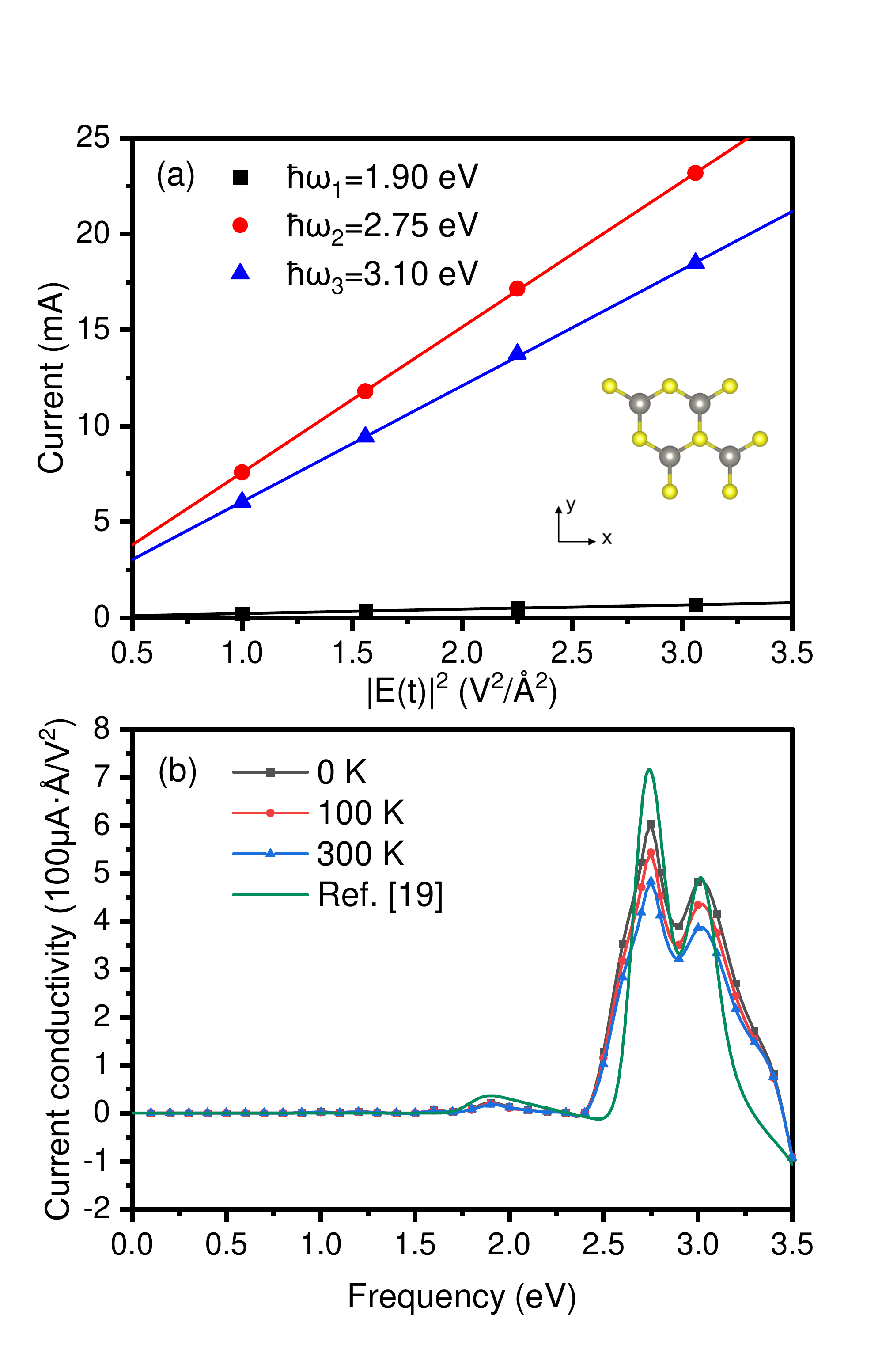}
  \caption{(a) The generated shift currents under different light intensity for $\hbar \omega$=1.9 eV, 2.75 eV, and 3.1 eV. The insert shows the structure of monolayer WS$_2$ .  (b)Shift current conductivities $\sigma^{yyy}$ as functions of excitation energy at 0 K, 100 K and 300 K, respectively.
  The conductivity calculated by the perturbation theory from Ref.~\cite{wang_first-principles_2019} is also shown as a comparison. }\label{fig:1}
\end{figure}

We first investigate the shift current conductivity $\sigma^{\alpha \beta \beta}(\omega)$ of monolayer WS$_{2}$\cite{wang_first-principles_2019}.
We calculate the shift-current at each frequency $\omega$ under different  excitation powers.
Figure~\ref{fig:1}(a) depicts the shift currents as functions of the excitation power for photon energies $\hbar \omega$=1.9 eV, 2.75
eV and 3.1 eV at zero temperature.
As we see, the shift currents increase linearly with the excitation power, and the shift current conductivity
can be fitted accordingly using the following relation\cite{sturman_photovoltaic_1992},
\begin{equation}
J^{\alpha}=\sigma^{\alpha \beta \beta}(\omega) E^{\beta}(\omega) E^{\beta}(-\omega)\, ,
\end{equation}
where $\alpha$ and $\beta$ are the lattice directions.

The fitted conductivity $\sigma^{yyy}(\omega)$ as a function of the excitation energy at $T$=0 K is shown in Fig.~\ref{fig:1}(b).
There are two major peaks: the first peak is at approximately 2.75 eV, with  $\sigma^{yyy}$ = 0.60 mA$\cdot$\AA/V$^{2}$,
and the second peak is at approximately 3.1 eV, with $\sigma^{yyy}$ = 0.48 mA$\cdot$\AA/V$^{2}$. There is also a very weak peak at approximately 1.9 eV.
The calculated current along the $x$-axis is zero conductivity due to the mirror symmetry $\hat{M}^y$, and therefore
$\sigma^{xxx}$ and $\sigma^{xyy}$  are zero \cite{wang_first-principles_2019,wang_first-principles_2017}.
The conductivities calculated at $T$=0 K in this work are in  good agreement with the previous results obtained by a second-order perturbation theory method \cite{wang_first-principles_2019,wang_first-principles_2017}, considering the results are from two very different approaches. The  conductivities calculated by rt-TDDFT are a little more smeared than those from perturbation theory, where the smearing is controlled by an artificial parameter.

Thus far, we have reproduced the shift-current conductivity of previous calculations. However, this method can go beyond the previous perturbation theory,
and allows us to calculate the shift-current at finite temperatures.
In these calculations, the dynamics of ions are described by the classical MD. The  shift-current conductivity  at 100 K and 300 K are also shown in  Fig.~\ref{fig:1}.
For the excitation energy $\hbar \omega$=2.75 eV, $\sigma^{yyy}$  decreases from 0.60 mA$\cdot$\AA/V$^{2}$  at $T$=0 K, to  0.54 mA$\cdot$\AA/V$^{2}$  at 100 K,
and 0.48 mA$\cdot$\AA/V$^{2}$  at 300 K.
Similarly, for $\hbar \omega$=3.1 eV, $\sigma^{yyy}$ decreases from 0.48 mA$\cdot$\AA/V$^{2}$  at $T$=0 K, to 0.44  mA$\cdot$\AA/V$^{2}$  at $T$=100 K, and 0.39 mA$\cdot$\AA/V$^{2}$ and 300 K.
The  shift-current conductivity gradually decreases with the increasing of temperature, presumably due to electron-phonon scattering.

To further analyze the contribution to the shift current,  we project the time-dependent wave functions $\tilde{\psi}_{n k}(t)$
to KS orbitals $\tilde{\psi}_{m k}^{0}$  of the ground state, i.e., the KS orbitals before applying the excitation light,
\begin{equation}
| \tilde{\psi}_{n \boldsymbol{k}}(t) \rangle=\sum_{m} c_{n m,\boldsymbol{k} }(t)  | \tilde{\psi}_{m\boldsymbol{k}}^{0}\rangle\, ,
\end{equation}
where $m$ is the band index of the ground state Kohn-Sham orbitals.
The shift current can then be written as,
\begin{equation}
\langle \tilde{\psi}_{n \boldsymbol{k}}(t)|\nabla|\tilde{\psi}_{n \boldsymbol{k}}(t) \rangle=\sum_{m, m^{\prime}} c_{n m,\boldsymbol{k}}^{*}(t) c_{n m^{\prime},\boldsymbol{k}}(t) \langle \tilde{\psi}_{m \boldsymbol{k}}^{0}|\nabla|\tilde{\psi}_{m^{\prime} \boldsymbol{k}}^{0}\rangle \, .
\label{eq:decomposition}
\end{equation}
We find that the shift currents calculated via Eq.\ref{eq:decomposition} are almost identical to the original results.
The major contribution to the shift currents in the studied energy range,
comes from the highest occupied bands and the lowest unoccupied bands, which are consistent with previous work \cite{wang_first-principles_2019}.

The total shift current can be decomposed into the off-diagonal contribution, i.e., the contribution from $m \neq m^{\prime}$ terms, and
the diagonal contribution, i.e.,  the $m = m^{\prime}$ terms.
We can further decompose the diagonal contribution from electrons (i.e., $m$ sum over the unoccupied bands) and holes (i.e., $m$
sum over the occupied bands). The diagonal terms are related to the quantity $\boldsymbol{\Delta}_{mn}(\boldsymbol{k})=\boldsymbol{v}_{mm}(\boldsymbol{k})-\boldsymbol{v}_{nn}(\boldsymbol{k})$,
where $\boldsymbol{v}_{mm}(\boldsymbol{k})$ and $\boldsymbol{v}_{nn}(\boldsymbol{k})$ are the velocities of the electrons and holes respectively
\cite{sipe_second-order_2000,ibanez-azpiroz_ab_2018}. $\boldsymbol{\Delta}_{mn}$ has a clear physical meaning that this part of the shift current arises from
the electron velocity change on the optical transition.

\begin{table}[!tbp]
\centering
  \caption{The total shift currents (mA) and the off-diagonal currents and the diagonal current in
  the single-layer WS$_{2}$.  The excitation power is $1.33 \times 10^{13} \mathrm{~W} / \mathrm{cm}^{2}$, at $T$=0 K.}
  \label{table:decompose}
  \begin{tabular}{c|c|c|c}
  \hline\hline
      \multirow{2}*{Frequency}&{Total current}&{Off-diagonal}&{Diagonal}\\
						{ } & (mA) & (mA) &(mA)\\
  \hline
  2.75 eV & 7.57 & 0.35 & 7.22 \\
  3.10 eV & 6.03 & 0.73 & 5.30 \\	
  \hline\hline				
  \end{tabular}
  \end{table}

Table \ref{table:decompose} lists the total shift currents and the off-diagonal and diagonal components for
 $\hbar \omega$=2.75 eV and 3.10 eV.
As we see from the Table, the shift-current is dominated by the dianoal contribution. For $\hbar \omega$=2.75 eV,
off-diagonal contribution account for only roughly 5\% of the total current, and the rest 95\% is contributed by the diagonal current.
Similar results are also obtained for the  $\hbar \omega$=3.10 eV case.
The diagonal current contributes 88\% of the total current, whereas the off-diagonal term
only contributes 12\%. Whether this is a general result for other materials remains to be further studied.

%\textbf{Excitation}

\begin{figure}
  \centering
  \includegraphics[width=0.5\textwidth]{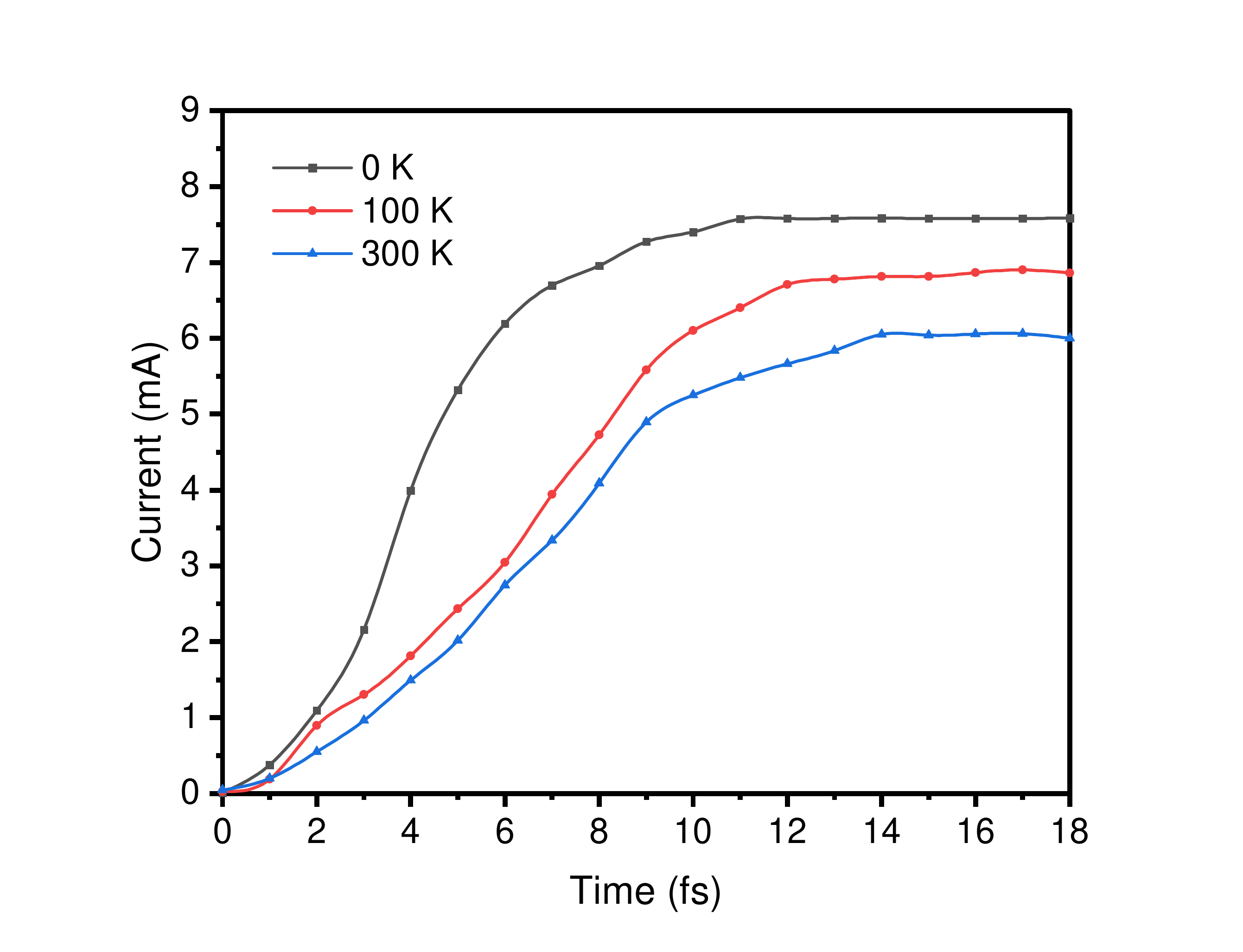}
  \caption{The generation of shift currents after turning on the light with the excitation of $1.0 \times 10^{10} \mathrm{~W} / \mathrm{cm}^{2}$ at 0 K, 100 K and 300 K, respectively.}\label{fig:2}
\end{figure}

One of the most prominent advantages of the rt-TDDFT method is that it can explore the ultrafast dynamics of the process.
It is interesting to see how fast the system can establish steady shift current under illumination.
 Figure~\ref{fig:2} shows the shift current as function of time after turning on the laser with $\hbar \omega$=2.75 eV.
 As one can see, the shift current quick raises after applying the laser, and the
 saturation times of the shift current after excitation are about
 11 fs, 12 fs and 14 fs at temperatures of 0 K, 100 K and 300 K, respectively.
This suggests that the shift-current  effect can be used as ultra-fast photon detectors, whereas the typical response time for
photodiode based photon detectors is about a few tens of picoseconds.
Figure~\ref{fig:2} reveals that at the same excitation power, the saturated shift currents at 100 K is 89\%  of that at 0 K,
whereas the saturated shift currents at 300 K is only about 80\% of that at 0 K.
This could be understood that at higher temperatures,
the electron-phonon scattering is stronger, therefore the generated shift current would be weaker.
Similar results are obtained for the $\hbar \omega$=3.10 eV.
%  (see Supporting Information).

\begin{figure}
  \centering
  \includegraphics[width=0.5\textwidth]{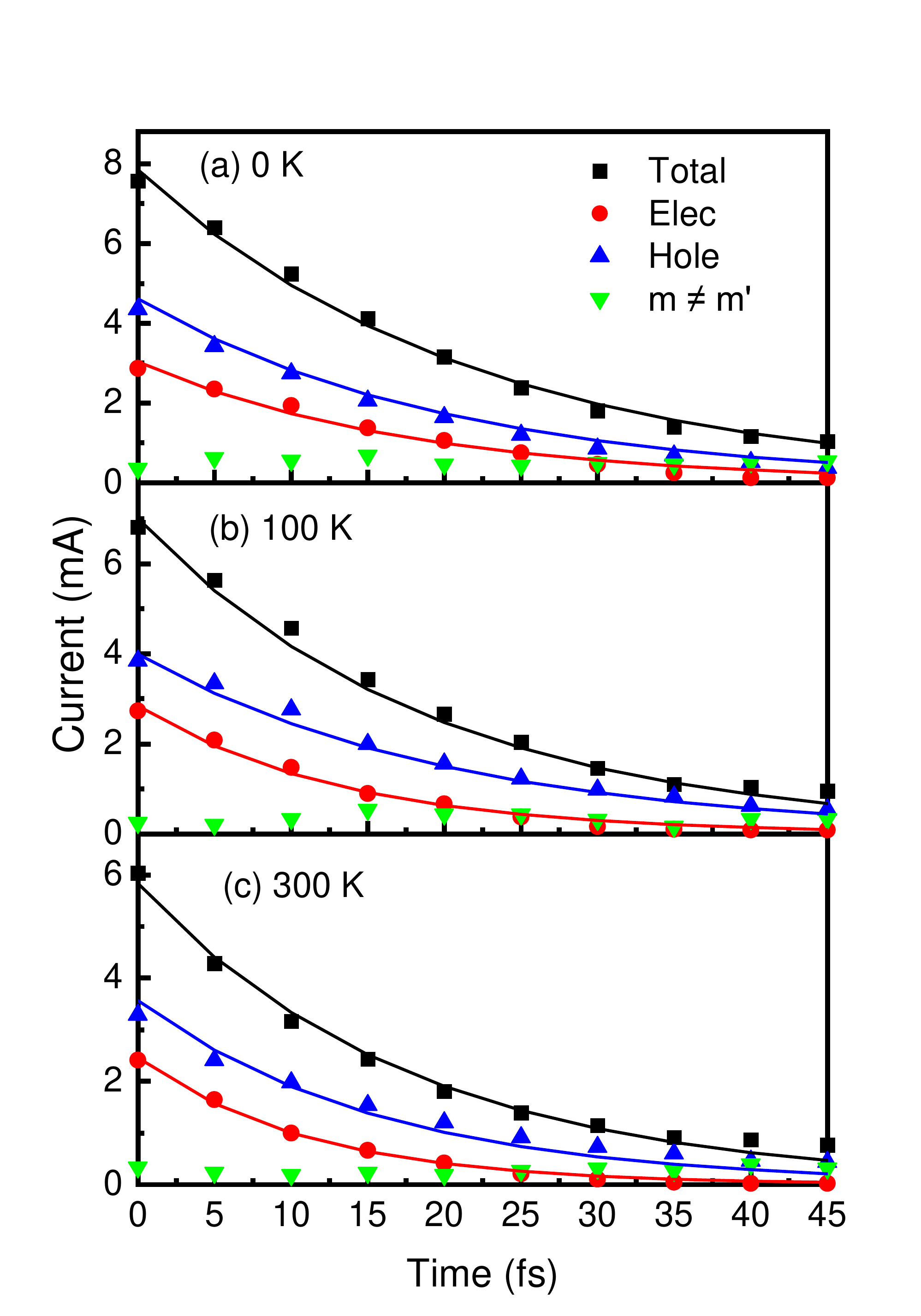}
  \caption{The dissipation of the total, electron, hole and off-diagonal ($m$$\neq$$m'$) currents at (a) 0 K, (b) 100 K and (c) 300K, respectively, after turning off the light field. }\label{fig:3}
\end{figure}

After turning off the lights, the shift current would dissipate.
The dissipation time of the shift current, which determines the diffusion length of the excited carriers, is crucial for the PV device  applications
\cite{spanier_power_2016}.
To calculate the current relaxation time, we turn off the light after the current becomes steady, and monitor the decay of the current.
Figure~\ref{fig:3}(a-c) depict the decay of the shift current after turning off the light at 0 K, 100 K and 300 K respectively at $\hbar \omega$=2.75 eV.
We show the results for the total currents, and the components including the electrons, holes, as well as the off-diagonal currents.
The total shift current, electronic current and hole current all decay (roughly) exponentially.
Specifically, the electron currents decrease  almost to zero at about 30 - 45 fs depending on the temperature. However, the total currents
and the hole currents have some (about 10 \% of the steady currents) long-lived residual components, which do not decay to zero in the simulated time range.
Remarkably, the off-diagonal current only has small oscillations  between 0.10 mA and 0.52 mA,  and does not decay in the simulation up to 45 fs for all temperatures.

We further fit the current relaxation times of the total, electron and hole currents at different temperatures via exponential functions,
and the results are shown in Table \ref{table:relaxationtime}.
The relaxation times of the shift currents decrease with increasing temperature, as expected, due to the stronger electron-phonon scattering.
For the total shift current, the  relaxation time decays from 21.84 fs at 0 K to 17.84 fs at 300 K. The relaxation times  of electrons (around 11 fs - 18 fs)
are a little shorter than those of holes (around 15 fs - 20 fs). The small long-life residuals of the total and hole currents cannot be fitted very well due to the limited simulation time. In fact, the carrier relaxation times are not very sensitive to temperature, which reflects the band topological nature of the shift current\cite{morimoto2016topological}.

We note that the relaxation times of the shift currents are not identical to the carrier lifetimes
due to electron-phonon scattering or electron-impurity scattering,
as the former ones are finite (and very short) even in perfect crystals at zero temperature.
The ultrafast shift current relaxation times are probably due to electron-electron scattering.
The diffusion length $l_0$, which can be estimated as $l_0\approx \tau \cdot v_{\bf k}$.
The very short current relaxation times are responsible for the extremely low BPV effects in bulk materials, because most photon generated
hot carriers are relaxed before they reach the electrode \cite{spanier_power_2016}. However, when the distance between the electrodes is comparable to
the diffusion length, the BPV effects are greatly enhanced \cite{alexe2011tip,zenkevich2014giant,spanier_power_2016,Li2021CuInP2S6}.
The mechanisms of the shift-current relaxation and the
calculation of the diffusion length are extremely important for the applications, which are beyond the scope of this work. We leave them for future studies.

\begin{table}[t!]
  \begin{center}
    \caption{Lifetimes of shift currents and their components at different temperatures with photo energy
    $\hbar \omega$=2.75 eV.}
    \label{table:relaxationtime}
    \begin{tabular}{c|c|c|c}
    \hline\hline
       Temperature  & Total current (fs) & Electron (fs) & Hole (fs) \\
      \hline
          0 K &  21.84 & 17.71 & 20.40 \\
     % \hline
      100 K &19.39 & 13.36 & 20.33 \\
   %   \hline
      300 K &17.84 & 11.08 & 15.85 \\
      \hline\hline
    \end{tabular}
  \end{center}
\end{table}

%\section{Summary}

To summarize, the dynamics of shift current are extremely important not only for fundamental science, but also  for  device applications.
We investigate the shift current dynamics in the  WS$_{2}$ monolayer by using the rt-TDDFT method.  We find that the shift-current can be established within 10 - 20 fs,
and the current dissipates within approximately a few tens of femtoseconds. This property can be used for ultrafast photon detection.
This work is an important step toward understanding the dynamics of shift-current effects.

%\acknowledgements
We thank Prof. Y. Xu for sharing their data of shift current conductivities with us. This work was funded by the Chinese National Science
Foundation Grant Numbers 11774327, 12134012, 11874335, and 12188101. The numerical calculations were performed on the USTC HPC facilities.

%\bibliographystyle{apsrev}
%\bibliography{WS2.bib,shiftcurrent.bib}


\end{document}